\documentclass[prb,twocolumn,showpacs,floatfix,nobalancelastpage]{revtex4}
\usepackage{bm}
\usepackage{dcolumn}
\usepackage{graphicx}

\begin{document}

\title{Maximally--localized Wannier Functions in Antiferromagnetic MnO
within the FLAPW Formalism}

\author{Michel Posternak}
\email{postma@dpmail.epfl.ch}
\author{Alfonso Baldereschi} 
\affiliation{Institute of Theoretical Physics, Faculty of Basic Sciences, Swiss
Federal Institute of Technology Lausanne, EPFL, PHB--Ecublens, CH--1015
Lausanne, Switzerland}
\author{Sandro Massidda} 
\affiliation{Istituto Nazionale di Fisica della Materia--Dipartimento di Fisica,
Universit\`a di Cagliari, Cittadella Universitaria, I-09042 Monserrato (CA),
Italy}
\author{Nicola Marzari}
\affiliation{Department of Materials Science and Engineering, Massachusetts
Institute of Technology, Cambridge, Massachusetts 02139--4307}

\date{\today}

\begin{abstract}
We have calculated the maximally--localized Wannier functions of MnO in its
antiferromagnetic (AFM) rhombohedral unit cell, which contains two formula
units. Electron Bloch functions are obtained with the
linearized--augmented--plane--wave method within both the LSD and the LSD+$U$
schemes. The thirteen uppermost occupied spin--up bands correspond in a pure
ionic scheme to  the five Mn 3$d$ orbitals at the Mn$_1$ (spin--up) site, and
the four O 2$s$/2$p$ orbitals at each of the O$_1$ and O$_2$ sites. Maximal
localization identifies uniquely four Wannier functions for each O, which are
trigonally--distorted $sp^3$--like orbitals. They display a weak covalent
bonding between O 2$s$/2$p$ states and minority--spin $d$ states of Mn$_2$,
which is absent in a fully ionic picture. This bonding is the fingerprint of
the interaction responsible for the AFM ordering, and its strength depends on
the one--electron scheme being used. The five Mn Wannier functions are centered
on the  Mn$_1$ site, and are atomic orbitals modified by the crystal field.
They are not uniquely defined by the criterion of maximal localization and we
choose them as the linear combinations which diagonalize the $r^2$ operator, so
that they display the D$_{\text{3d}}$ symmetry of the Mn$_1$ site.
\end{abstract}

\pacs{71.15.-m,71.15.Ap,75.30.Et,75.50.Ee}

\maketitle

\section{INTRODUCTION}
\label{sec:intro}

The mean--field one--particle description of the electronic struture of 
periodic crystalline solids is usually based on extended Bloch functions (BFs).
Within the Born--von Karman periodic boundary conditions, the cyclic
translational subgroup, which commutes with the effective one--electron
Hamiltonian, contains $N$ translations by corresponding direct lattice vectors
$\mathbf{R}$. This Abelian subgroup has $N$ one--dimensional irreducible
representations, which are labeled with wavevectors  $\mathbf{k}$ within the
first Brillouin zone (BZ). Therefore, extended Bloch states
$\psi_{m\:\mathbf{k}}(\mathbf{r})$ in this description are classified with two
quantum numbers, the  band index $m$ and the crystal--momentum $\mathbf{k}$,
and are obtained by diagonalization of the effective  one--electron
Hamiltonian. An alternative description can be derived in terms of localized
Wannier functions~\cite{Wannier37} (WFs) $w_n (\mathbf{r}-\mathbf{R})$, which
are defined in real space via a unitary transformation performed on the Bloch
functions. They are also labeled with two quantum numbers, the orbital index
$n$ and the direct lattice vector $\mathbf{R}$ indicating the unit cell they
belong to. In contrast to BFs, WFs are useful, for example, in visualizing
chemical bonds or in describing the dielectric properties of non--metallic
materials. They can be considered as the generalization to solids of the
concept of ``localized molecular orbitals'' for finite systems~\cite{Boys96}.
However, one major problem in practical calculations within this representation
is the non--uniqueness of WFs, related to the phase arbitrariness of the BFs,
and to the arbitrary unitary transformations which can be performed on the BFs
at any given $\mathbf{k}$--point in the BZ. As demonstrated recently by Marzari
and Vanderbilt~\cite{Marzari97}, this non--uniqueness can be resolved in
principle by imposing the further condition of maximal localization. However,
some residual arbitrariness remains, related to the choice of the localization
criterion.

The late transition--metal (TM) monoxides MnO, FeO, CoO, and NiO challenge the
theory of electronic states since several decades. These highly correlated
materials feature a Mott insulator character, and the conventional
local--spin--density (LSD) scheme~\cite{Terakura84} gets into difficulties with
the localized TM $d$--orbitals, predicting in particular incorrectly their
spectral weight and their energy relative to O~$2p$ states. All these oxides
have the  rocksalt structure in their paramagnetic phase, while below the Neel
temperature $T_{\text{N}}$, a type II antiferromagnetic (AFM) ordering
occurs~\cite{Roth58}. The experimentally well--documented compound MnO, which
is considered to be in the  intermediate charge--transfer/Mott--Hubbard regime,
is a particularly suitable  case study, since the available theoretical schemes
apply best to this material. Indeed, because of the exchange stabilization of
its half--filled $d$ shell, even LSD predicts its insulating character,
although the corresponding energy gap and magnetic moment are much smaller than
experimental data. Several electronic structure calculations have been
performed for MnO, using traditional band--structure schemes, like
unrestricted  Hartree--Fock (UHF)~\cite{Massidda99} and local--spin density
(LSD), as well as more innovative approaches taking into account at various
levels the large value of the on--site  Coulomb repulsion for the metal 3$d$
states, like the SIC~\cite{Svane90} and the LSD + $U$~\cite{Anisimov91}
methods, and the more recent model $GW$ scheme~\cite{Massidda95,Massidda97}.
Among the quantities calculated within the latter approach, special attention
has been devoted to the quasi--particle  spectrum~\cite{Massidda95}, and
recently to the zone--center optic phonon frequencies and the Born effective
charge tensor~\cite{Massidda99}. Comparison with experimental data of the
various physical quantities computed in the different schemes, has demonstrated
a monotonic trend with the separation energy between occupied and empty $d$
bands, which is too small within LSD, too high within UHF, and has about the
correct value within model $GW$. This energy separation is strongly related to
the on--site interaction $U$. The analysis of all these results, has been
performed in terms of the extended Bloch states. 

On the other hand, no \textit{ab--initio} investigation exists of the
electronic states in AFM MnO in terms of localized Wannier functions. Keeping
in mind the inherent limitations of the mean--field one--particle approaches,
it would be however instructive to materialize within the WFs description how
superexchange (which is responsible for AFM ordering) manifests itself in a
first--principles one--electron picture. Furthermore, the maximal localization
method of Marzari and Vanderbilt~\footnote{A generalization of the method of
Ref.~\onlinecite{Marzari97} for the case of entangled energy bands has been
given by I. Souza, N. Marzari, and D. Vanderbilt, Phys. Rev. B \textbf{65},
035109 (2001).} has been applied in the past to several periodic
systems~\cite{Marzari97,Marzari98,Souza00}, but not yet to a low (trigonal)
symmetry solid like AFM MnO, nor to a compound with partially filled cation $d$
shell.

Motivated by the above reasons, we present in this work an \textit{ab--initio}
calculation of the maximally--localized Wannier functions of AFM MnO
corresponding to the uppermost occupied bands, using the all--electron,
full--potential linearized--augmented--plane--wave (FLAPW)
method~\cite{Jansen84,Massidda93}. Because the trends in the electronic
properties from UHF to LSD are already known for this system, most of the
computations have been performed for convenience within the latter scheme. To
investigate the effects related to the on--site Coulomb interaction, we have
also used the LSD + $U$ method, implemented with the same FLAPW technical
ingredients~\cite{Shick99} used in the LSD computations.

The manuscript is organized as follows. In Sec.~\ref{sec:method} we give
technical details of the FLAPW implementation, and structural informations
relevant to AFM MnO. The necessary ingredients of our WFs calculations are then
introduced. As we follow closely the formalism of Ref.\ \onlinecite{Marzari97},
we will not describe in detail the method itself. A general result regarding
maximal localization of $s$--$p$ WFs is deferred to an Appendix. In
Sec.~\ref{sec:results}, we present and discuss the results of our
maximally--localized WFs calculations. Finally, in Sec.~\ref{sec:concl} we draw
our conclusions.

\section{METHOD}
\label{sec:method}

\subsection*{A. FLAPW calculations}
\label{subsec:su_lapw}

All quantities presented and discussed in this paper have been computed using
the semirelativistic FLAPW method~\cite{Jansen84,Massidda93}, expanded with
local orbitals~\cite{Singh91} (lo) where appropriate. Inclusion of lo's in
addition to the normal FLAPW basis enforces mutual state orthogonality and
increases variational freedom. This allows to treat the semicore Mn~$3s$, $3p$
states together with the valence states, and helps in dealing with the
linearization of Mn~$3d$ and O~$2s$, $2p$ states. Core states are calculated
fully relativistically and self--consistently in the crystal potential. For the
LSD calculations, the Hedin and Lundqvist exchange--correlation functional has
been adopted. In the calculations performed within the LSD + $U$ method, the
values of the Hubbard and exchange constants, $U = 6.9$ eV and $J = 0.86$ eV,
have been taken from Ref.\ \onlinecite{Anisimov91}. The atomic--sphere radii
for Mn and O are chosen to be 2.0 and 1.8 a.u., respectively, and the FLAPW
basis size is set to include all plane waves with energy up to 16.0 Ry. Four
special points inside the irreducible wedge of the BZ were used for evaluating
the charge density during self--consistency cycles.

\subsection*{B. Structural details}
\label{subsec:su_stru}

\begin{figure}[htbp!]
\includegraphics[scale=0.5]{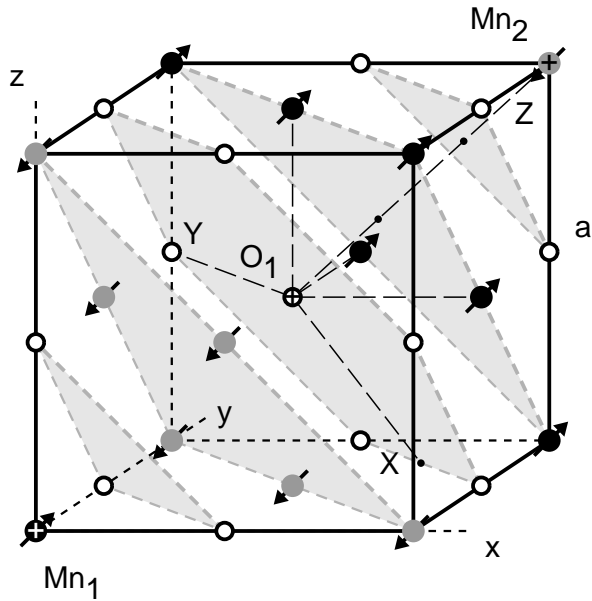}
\caption{Antiferromagnetic structure of MnO. Magnetic moments are arrayed in
ferromagnetic sheets (shaded areas) parallel to (111) planes. Mn$_1$, Mn$_2$,
and O$_{1,2}$ sites are given by large black, large grey, and small open
circles, respectively. The trigonal coordinate axes are also indicated with
long--dashed lines.\label{fig1}}
\end{figure}

AFM ordering occurs in MnO below $T_{\text{N}} = 117$~K along any one of the
$\langle 111 \rangle$ directions. The transition is accompanied by a small
crystallographic distortion that transforms the cubic structure into a
rhombohedral one~\cite{Landolt84}: the $90^{\circ}$ angle between the lattice
vectors increases only by $0.624^{\circ}$. We have shown~\cite{Massidda99}
earlier that the effect of this distortion on the calculated zone--center optic
phonon frequencies and Born effective charge tensor is negligible compared to
the anisotropy induced solely by the magnetic order. Therefore, throughout the
present study we use the ionic positions of the perfect rocksalt chemical cell,
with the experimental lattice constant value $a = 4.435$ \AA. In the AFM phase,
the magnetic moments of MnO are arranged into ferromagnetic sheets which are
parallel to (111), while the direction of magnetization in neighboring planes
is reversed. The magnetic and crystalline structure of MnO is shown in
Fig.~\ref{fig1}. The rhombohedral magnetic cell, whose volume is twice that of
the paramagnetic rocksalt one, corresponds to the space group D$^5_{3\text{d}}
\;(R\overline{3}m)$. Several choices for the primitive translation vectors are
possible. However, in order to minimize discrete mesh effects in
$\mathbf{k}$--space integrations, and also for taking advantage of symmetry
properties (see below), we consider here the following primitive translation
vectors (in the cubic coordinate system):
$\mathbf{t}_1= a\:(1,\:\frac{1}{2},\:\frac{1}{2})$, 
$\mathbf{t}_2= a\:(\frac{1}{2},\:1,\:\frac{1}{2})$, and
$\mathbf{t}_3= a\:(\frac{1}{2},\:\frac{1}{2},\:1)$. The two equivalent anions
O$_1$ and O$_2$, whose site symmetry is C$_{3\text{v}}$, are located at
positions $\pm a\:(\frac{1}{2},\:\frac{1}{2},\:\frac{1}{2})$, while the two
non--equivalent cations Mn$_1$ (spin--up) and Mn$_2$, whose site symmetry is
D$_{3\text{d}}$, are at $(0,\:0,\:0)$ and $a\:(1,\:1,\:1)$. With our choice of
origin on Mn$_1$, the two Oxygen sites are therefore related by spatial
inversion.

\subsection*{C. Maximally localized Wannier functions}
\label{subsec:su_wann}

We consider the general case of a \textit{composite group} of energy bands
labeled by their band index $m=1,2,\ldots,f$, and connected among themselves by
degeneracies along high symmetry lines, but isolated from all other bands at
lower and higher energy. The phase arbitrariness (gauge dependence) of each $|
\psi_{m\mathbf{k}} \rangle$ with respect to the set of allowed $\mathbf{k}$
values propagates to the corresponding Wannier functions $| w_{n\mathbf{R}}
\rangle$, labeled by $n=1,2,\ldots,f$ and lattice vectors $\mathbf{R}$. This
resulting non--uniqueness manifests itself trough an arbitrary unitary matrix
$U_{mn}^{(\mathbf{k})}$ appearing in the most general transformation between
BFs and WFs, which is given by
\begin{equation}
| w_{n\mathbf{R}} \rangle = 
\frac{V}{(2\pi)^3} \int_{BZ} d\mathbf{k} \:e^{-i\mathbf{k} \cdot
\mathbf{R}} \sum_m U_{mn}^{(\mathbf{k})} \: | \psi_{m\mathbf{k}} \rangle,
\label{WF_i}
\end{equation}
where $V$ is the real--space primitive cell volume. A direct consequence of
this arbitrariness is that linear combinations of the WFs in the set $\{ |
w_{n\mathbf{R}} \rangle \}$ also form a suitable basis.

The strategy of Ref.\ \onlinecite{Marzari97} is to pick out from this arbitrary
choice of WFs, the particular set which is maximally localized according to
some criterion. Once this criterion has been chosen, and the composite group of
bands specified, the search for the set of maximally localized WFs becomes a
problem of functional minimization in the space of the matrices $U_{mn}^{\bf
(k)}$. The selected functional, which measures the sum of the quadratic spreads
of the WFs probability density, is given by
\begin{equation}
\Omega = \sum_n [\langle r^2 \rangle_n - \langle \mathbf{r} \rangle^2_n],
\label{omega_def}
\end{equation}
where $\langle \mathbf{r} \rangle_n = \langle w_{n\mathbf{0}}| \mathbf{r}
|w_{n\mathbf{0}} \rangle = i\:V/(2\pi)^3 \int_{BZ} d\mathbf{k} \:\langle
u_{n,\mathbf{k}}|\bm{\nabla}_{\mathbf{k}}| u_{n,\mathbf{k}} \rangle$ and
$\langle r^2 \rangle_n = \langle w_{n\mathbf{0}}| r^2 |w_{n\mathbf{0}}
\rangle_n$. In these expressions, $u_{n,\mathbf{k}}\:(\mathbf{r}) = 
e^{-i\mathbf{k} \cdot \mathbf{r}} \sum_m U_{mn}^{(\mathbf{k})} \:
\psi_{m\mathbf{k}} \: (\mathbf{r})$ is a periodic function which can be
obtained from the Bloch functions of the composite group of bands. Practically,
one calculates the BFs on equispaced Monkhorst--Pack~\cite{Monkhorst76} meshes
of $\mathbf{k}$ points in the unit cell $\overline{\text{BZ}}$ (whose volume is
equal to the conventional BZ one), built on the reciprocal lattice vectors. The
grids have been offset in order to include $\Gamma$. With our chosen cell
geometry of MnO, the reciprocal lattice vectors are:
$\mathbf{g}_1 = \pi/a\: (3,\:-1,\:-1)$,
$\mathbf{g}_2 = \pi/a\: (-1,\:3,\:-1)$, and
$\mathbf{g}_3 = \pi/a\: (-1,\:-1,\:3)$.
We then express the matrix elements of the gradient $\mathbf{\nabla_k}$
appearing in the localization functional in terms of finite differences. As
shown in  Ref.\ \onlinecite{Marzari97}, the only information needed is the
overlap matrix $M_{nn'}^{(\mathbf{k},\mathbf{b})} = \langle u_{n\mathbf{k}} |
u_{n',\mathbf{k}+\mathbf{b}} \rangle$, where $\mathbf{b}$ are vectors
connecting each mesh point to its nearest neighbors. From these quantities, the
gradient of the spread functional $\Omega$  with respect to an infinitesimal
unitary transformation $\delta U_{mn}^{(\mathbf{k})}$ of the BFs can also be
evaluated. Once this gradient is computed, the minimization can take place via
a steepest descent or conjugate--gradient algorithm.

It is important to stress that if the minimum of $\Omega$ is flat (as it will
be shown to be the case for some of the MnO WFs), its precise location may be
hindered by the various numerical approximations involved in the calculations.
In this work, we have especially taken care of the following features, which
clearly become irrelevant in the limit of a dense mesh ($N \rightarrow \infty$,
$b \rightarrow 0$): (i) the regular mesh $\{ \mathbf{k} \}_{\:\overline{BZ}}$
has been chosen in order to have the lattice symmetry, \textit{i.e.} it
transforms into itself by application of the point group operations $R$ of the
crystal; (ii) the shell of $\mathbf{b}$ vectors used in the finite--difference
formula for $\bm{\nabla}_{\mathbf{k}}$ has been constructed on the 12 vectors
of type $\{3\:3\:2\}$ and $\{1\:\bar{1}\:0\}$ (in internal units). They are the
sides midpoints of cubes which can be built on particular points of the
rhombohedral mesh, and this set of 12 points is therefore compatible with the
highest possible symmetry.

Finally, and following the procedure described in Eqs. (62)--(64) of Ref.\
\onlinecite{Marzari97}, we start the minimization procedure by constructing a
set of ``trial functions'' in the unit cell, which are an initial guess of the
final WFs. We use Gaussians centered on atomic sites and modulated by an
appropriate combination of spherical harmonics, with a rms width value such
that the gaussian is negligible outside the corresponding atomic sphere. A
unitary rotation among the initial BFs is then performed in order to maximize
their projection on the trial functions.

It has been shown~\cite{Krueger72} that Wannier functions can be chosen with
well--defined symmetry properties, and forming (generally reducible)
representations of the crystal point group. In order to investigate the nature
and possible reductions of these representations in the case of MnO, and also
to be in the position to use group--theory methods for further analysis, we
need to calculate the representation matrix elements $\langle w_{n'\mathbf{0}}
| P_R \: | w_{n\mathbf{0}} \rangle$, where $P_R$ is the transformation operator
corresponding to the operation $R$ of the point group, and the WFs are in the
central cell. For simplicity, we restrict the formalism to symmorphic space
groups (which is actually the case for D$^5_{3\text{d}}$). Using
Eq.~(\ref{WF_i}) with the $\mathbf{k}$ points discretization and
$\mathbf{R}=\mathbf{0}$, we have for the rotated Wannier function
\begin{equation}
P_R \: | w_{n\mathbf{0}} \rangle = \frac{1}{N} \sum_{\mathbf{k} \in
\overline{BZ}} \sum_m U_{mn}^{(\mathbf{k})} \: P_R \: | \psi_{m\mathbf{k}}
\rangle,
\end{equation}
and, with the ``periodic gauge'' condition $\psi_{m\:R\mathbf{k}}\:
(\mathbf{r}) = \psi_{m\:R\mathbf{k} + \mathbf{G_k}}\:(\mathbf{r})$
\begin{widetext}
\begin{eqnarray}
\langle w_{n'\mathbf{0}} | P_R \: | w_{n\mathbf{0}} \rangle_{NV} &=& \frac{1}
{N^2} \sum_{\stackrel{\mathbf{k},\mathbf{k'}} {\in \overline{BZ}}} \sum_{m,m'}
U_{m'n'} ^{*(\mathbf{k'})} \:U_{mn}^{(\mathbf{k})} \: \langle
\psi_{m'\mathbf{k'}} | P_R | \psi_{m\mathbf{k}} \rangle_{NV} \nonumber \\*
&=& \frac{1}{N} \sum_{\mathbf{k} \in \overline{BZ}} \sum_{m,m'}
U_{m'n'}^{*(R\mathbf{k}+\mathbf{G_k})} \: U_{mn}^{(\mathbf{k})} \: \langle
\psi_{m'\:R\mathbf{k}+\mathbf{G_k}} | P_R | \psi_{m\mathbf{k}} \rangle_V.
\label{warep}
\end{eqnarray}
\end{widetext}
The last braket is trivial only in the case of non--degenerate BFs. In the
general case, on the other hand, we have
\begin{eqnarray}
P_R \: \psi_{m\mathbf{k}}\:(\mathbf{r}) &=&
\psi_{m\mathbf{k}}\:(R^{-1}\mathbf{r}) \nonumber \\*
&=& \sum_{m'} D_{m'm}(R,\:\mathbf{k}) \:\psi_{m'\:R\mathbf{k}}\:(\mathbf{r}), 
\label{bloch}
\end{eqnarray}
where $D_{m'm}(R,\:\mathbf{k})$ is a unitary transformation associated with the
symmetry operation $R$, and the summation is over states degenerate with
$\psi_{m'\:R\mathbf{k}}\:(\mathbf{r})$. If this latter state is
non--degenerate, $D_{m'm}(R,\:\mathbf{k})=\delta_{m'm}$, and Eq.~(\ref{bloch})
becomes the usual formula. We note that the above formulas allow to decompose
$|w_{n\mathbf{0}} \rangle$ into a linear combination of basis functions
transforming according to the relevant irreducible representations of the point
group.

\section{RESULTS}
\label{sec:results}

For our Wannier--function description of the electron states in AFM MnO, we
have considered as a single composite group the topmost thirteen valence bands
(O~$2s$, O~$2p$, Mn~$3d$). In this way, the corresponding Wannier functions
display the highest localization, and therefore can be considered as the
``elementary building blocks'' of the occupied electron states. As it is
convenient to interpret some of our results in terms of combinations of atomic
orbitals, we give in Table~\ref{table1} the relevant O$_{\text{h}}$
representations corresponding to $s$, $p$, and $d$ orbitals and their reduction
into irreducible representations of D$_{3\text{d}}$ and C$_{3\text{v}}$,
together with their corresponding basis. Cartesian coordinates in the parent
cubic reference system are written using $x$, $y$ and $z$ symbols, while $X$,
$Y$ and $Z$ correspond to the trigonal reference system ($Z$ axis along the
cubic [111] direction). Throughout this work, we use the Bethe notation for the
irreducible representations~\cite{Koster63}.
\begin{table}[htbp!]
\caption{Classification of $s$, $p$, and $d$ atomic orbitals according to the
irreducible representations of O$_{\text{h}}$ together with their reduction
into irreducible representations of D$_{3\text{d}}$ and C$_{3\text{v}}$, and
the corresponding basis. Bethe notation~\cite{Koster63} is used. The cubic and
trigonal coordinate systems are denoted by $x,\:y,\:z$ and $X,\:Y,\:Z$,
respectively.\label{table1}}
\begin{ruledtabular}
\begin{tabular}{l|l|l|l}
\multicolumn{2}{c|}{O$_{\text{h}}$ irreducible} &
\multicolumn{2}{c}{Reduction into D$_{3\text{d}}$ (C$_{3\text{v}}$)
irreducible} \\
\multicolumn{2}{c|}{representations} &
\multicolumn{2}{c}{representations and basis} \\
\hline
$\Gamma^+_1$ & 
$\begin{array}{l} \; s \end{array}$ & 
$\begin{array}{l} \Gamma^+_1 \;(\Gamma_1) \end{array}$ &  
$\begin{array}{l} \; s \end{array}$ \\
\hline
$\Gamma^-_4$ & 
$\left\{ \begin{array}{@{}l@{}} p_x \\ p_y \\ p_z \end{array} \right. $ &
$\begin{array}{l} \Gamma^-_2 \;(\Gamma_1) \\ \\ \Gamma^-_3 \;(\Gamma_3) \\
\end{array}$ &
$\begin{array}{l} \begin{array}{@{}l@{}} \; (p_x + p_y + p_z)/\sqrt{3} \; 
\equiv p_Z \end{array} \\ \left\{ \begin{array}{@{}l@{}}
(p_x + p_y -2p_z)/\sqrt{6} \; \equiv p_X \\
(p_x - p_y)/\sqrt{2} \; \equiv p_Y \end{array} \right. \end{array}$ \\
\hline
$\Gamma^+_3$ & $\left\{ \begin{array}{@{}l@{}} d_{x^2-y^2} \\ d_{z^2}
\end{array} \right. $ &
$\begin{array}{l}
\Gamma^+_3 \;(\Gamma_3) \end{array}$ & $\begin{array}{l} \left\{ 
\begin{array}{@{}l@{}} d_{x^2-y^2} \\ d_{z^2} \end{array} \right.
\end{array}$ \\
\hline
$\Gamma^+_5$ & 
$\left\{ \begin{array}{@{}l@{}} d_{xy} \\ d_{yz} \\ d_{zx} \end{array} 
\right. $ &
$\begin{array}{l} \Gamma^+_1 \;(\Gamma_1) \\ \\ \Gamma^+_3 \;(\Gamma_3) \\
\end{array}$ & 
$\begin{array}{l} \left. \begin{array}{@{}l@{}} \;
(d_{yz} + d_{zx} + d_{xy})/\sqrt{3} \; \equiv d_{Z^2} \end{array} \right. \\ 
\left\{ \begin{array}{@{}l@{}} (d_{yz} + d_{zx} - 2d_{xy})/\sqrt{6} \\
(d_{yz} - d_{zx})/\sqrt{2} \end{array} \right. \end{array}$ \\
\end{tabular}
\end{ruledtabular}
\end{table}

We have first studied the convergence of the spread functional $\Omega$ in
terms of the mesh density, using a sampling of the Brillouin zone
$\overline{\text{BZ}}$ defined by the Monkhorst--Pack meshes $\nu \times \nu
\times \nu$ with $\nu = 2$, 4, 6, and 8. Analysis has been done on the various
terms of the localization functional $\Omega$ discussed in  Ref.\
\onlinecite{Marzari97}. We found that a satisfactory level of convergence is
achieved using the 6$\times$6$\times$6 mesh, which represents a good compromise
between accuracy and computer burden. Also, such a mesh is large enough to
prevent difficulties related to finite grids and occurence of periodic WFs
replica. The 6$\times$6$\times$6 sampling has been retained throughout this
work.

The minimization of the localization functional, which determines the unitary
matrix $U_{mn}^{(\mathbf{k})}$, has been performed using a mixed strategy: a
simple fixed--step steepest--descent procedure is used during the first
iterations, followed by several iterations with a conjugate--gradient
procedure which is reset to steepest descent every 100 iterations. Because of
the delicate convergence of the Mn~WFs, and in order to be on the safe side, we
used a very large number (40000$-$80000) of iterations.
\begin{figure}[htbp!]
\includegraphics[scale=0.40]{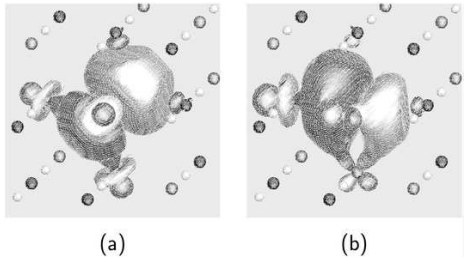}
\caption{LSD Wannier functions of AFM MnO corresponding to the whole spin--up
valence--band complex (13 bands). (a) The $2s/2p$ WF of O$_1$ with center along
the [111] direction; (b) one of the 3 equivalent O$_1$ WFs, with center close
to the [100] axis. The Mn$_1$, Mn$_2$, and O$_{1,2}$ sites are denoted by large
black, large grey, and small white spheres, respectively. Light grey and dark
grey colours indicate the positive and negative amplitudes of the WFs
contours.\label{fig2}}
\end{figure}

Several choices of trial orbitals have been tested to initialize efficiently
the minimization procedure. In particular (see Section~\ref{subsec:su_wann}),
we have considered the following sets of initial trial orbitals: (i) Gaussians
times real combinations of spherical harmonics $s$, $p_x$, $p_y$, and $p_z$ on
the O$_1$ and O$_2$ Oxygen sites, and $d_{z^2}$, $d_{x^2-y^2}$, $d_{xy}$,
$d_{yz}$, and $d_{zx}$ on the Manganese Mn$_1$ site. These real harmonics,
reported in Table~\ref{table1}, form the basis of reducible representations of
C$_{3\text{v}}$ (O site) and of D$_{3\text{d}}$ (Mn$_1$ site), and are oriented
along cubic axes; (ii) same centers as above, but with Gaussians modulated by
spherical harmonics combinations $s$, $p_X$, $p_Y$, $p_Z$, and $d_{Z^2}$,
$d_{X^2-Y^2}$, $d_{XY}$, $d_{YZ}$, and $d_{ZX}$, \textit{i.e.} basis functions
of the irreducible representations of C$_{3\text{v}}$ and D$_{3\text{d}}$;
(iii) combinations of the two above possibilities. All these sets of initial
trial orbitals eventually lead, at convergence, to the same WFs, which are
real, in agreement with the discussion in Ref.\ \onlinecite{Marzari97}.
However, the set (i) corresponds to a much faster convergence than the other
choices. The reason is directly related to the shape and orientation of the
final O~$2s$/$2p$ WFs, and this point will be discussed further below.

We give in Table~\ref{table2} the converged Wannier centers and spreads
corresponding to the four Oxygen WFs at the O$_1$ site for the spin--up chanel.
\begin{table}[htbp!]
\caption{Center and spread of the four Wannier functions for the O$_1$ site
corresponding to the uppermost LSD spin--up valence--bands of MnO. Center
components $\overline{r}_x$, $\overline{r}_y$, and $\overline{r}_z$ are given
with respect to the atomic site in the cubic reference system. Sum of WFs
centers and the total spread are also displayed. Values in parentheses
correspond to the LSD + $U$ scheme. Considering the C$_{3\text{v}}$ symmetry of
the O$_1$ site, WF 1 belongs to $\Gamma_1$ while WFs 2, 3, and 4 are equivalent
to each other. WFs 5--8 for the O$_2$ site have the same $\Omega$ and opposite
$\overline{\mathbf{r}}$.\label{table2}}
\begin{ruledtabular}
\begin{tabular}{cdddd}
\multicolumn{1}{c}{WF} &
\multicolumn{1}{r}{$\overline{r}_x$ (\AA)} &
\multicolumn{1}{r}{$\overline{r}_y$} (\AA) &
\multicolumn{1}{r}{$\overline{r}_z$} (\AA) &
\multicolumn{1}{r}{$\Omega$ (\AA$^2$)} \\
\hline
 1     &    0.174  &   0.174  &   0.174  &    0.751 \\
       &   (0.186) &  (0.186) &  (0.186) &   (0.686) \\
 2     &   -0.387  &   0.036  &    0.036 &    0.773 \\
       &  (-0.366) &  (0.049) &  (0.049) &   (0.704) \\
 3     &    0.036  &  -0.387  &   0.036  &    0.773 \\
       &   (0.049) & (-0.366) &  (0.049) &   (0.704) \\
 4     &    0.036  &   0.036  &  -0.387  &    0.773 \\
       &   (0.049) &  (0.049) & (-0.366) &   (0.704) \\
\hline
 Total &   -0.141  &  -0.141  &  -0.141   &   3.069 \\
       &  (-0.082) & (-0.082) & (-0.082)  &  (2.797) \\
\end{tabular}
\end{ruledtabular}
\end{table}
$\overline{r}_x$, $\overline{r}_y$, and $\overline{r}_z$ are the components of
the Wannier center position with respect to the corresponding atomic site. We
give also the total spread $\Omega$ of these four WFs, and the sum of the
corresponding centers whose interpretation will be discussed below. WFs related
to the O$_2$ site have the same $\Omega$ and opposite $\overline{\mathbf{r}}$.
The spreads corresponding to the five LSD WFs (9--13) at the Mn$_1$ site are
given in the first line of Table~\ref{table3}. 
\begin{table}[htbp!]
\caption{The spreads (in \AA$^2$) of the five Wannier functions (9--13)
centered at the origin (Mn$_1$ site) corresponding to the uppermost LSD and LSD
+ $U$ spin--up valence--bands of MnO. Total spreads are also indicated. The
meaning of s--LSD and s--LSD + $U$ is explained in the text.\label{table3}}
\begin{ruledtabular}
\begin{tabular}{lcccccc}
\multicolumn{1}{c}{ }  &
\multicolumn{1}{c}{9}  &
\multicolumn{1}{c}{10} &
\multicolumn{1}{c}{11} &
\multicolumn{1}{c}{12} &
\multicolumn{1}{c}{13} &
\multicolumn{1}{c}{Tot (9--13)} \\
\hline
LSD          &  0.6220 &  0.6218 &  0.6497 &  0.6493 &  0.6493 &  3.193 \\
s--LSD       &  0.5950 &  0.5950 &  0.6248 &  0.6899 &  0.6899 &  3.194 \\
s--LSD+$U$   &  0.4713 &  0.4713 &  0.5133 &  0.5163 &  0.5163 &  2.488 \\
\end{tabular}
\end{ruledtabular}
\end{table}
By symmetry, these WFs are centered on the Mn$_1$ site (origin). Because the
two Oxygen sites are equivalent, corresponding results for the spin--down
chanel can be simply obtained by reversing Mn$_1$ and Mn$_2$. The total spread
(sum of the 13 individual spreads) is 9.332 \AA$^2$. We note also that the sum
of centers has all cartesian components equal to zero. Indeed, this quantity
represents the electronic polarization (modulo a lattice vector) corresponding
to valence states, and has to be zero as the system is centrosymmetric. In
order to analyze in detail Table~\ref{table2}, it is useful to construct the
matrix elements of the representation of the symmetry operations of the
D$_{3\text{d}}$ point group on the basis of the 13 Wannier functions. This is
done by using Eq.~(\ref{warep}). First, we have verified by constructing the
multiplication table that these 12 matrices form indeed a representation of
D$_{3\text{d}}$. For non--trivial operations of the point group
D$_{3\text{d}}$, they display the following structure: (i) there is an
8$\times$8 block corresponding to O~$2s$/$2p$ Wannier functions. Each line of
this block consists of zeros, except for one element which is equal to 1.
Because the origin is on the Mn$_1$ site, the operations of D$_{3\text{d}}$ are
simply transforming all WFs on \textit{both} O sites into each other; (ii)
there is a 5$\times$5 block corresponding to Mn$_1$~$3d$ Wannier functions.
This block does not display a well--defined structure, and in general all its
elements are non--zero; (iii) the two off--diagonal blocks (connecting
O~$2s$/$2p$ and Mn$_1$~$3d$ WFs) have all their elements equal to zero, at an
accuracy better than $10^{-7}$. Therefore, the 13$\times$13 representation is
block--diagonal, and in particular the 5$\times$5 Mn$_1$~$3d$ block makes a
(reducible) representation by itself.

We discuss first the O~$2s$/$2p$ Wannier functions. The two non--equivalent WFs
corresponding to site O$_1$ are shown in Fig.~\ref{fig2}, with a viewpoint
which is about the same as in Fig.~\ref{fig1}. The equivalent WFs corresponding
to site O$_2$ can simply be obtained by inversion symmetry with respect to the
origin. In the absence of discrimination between Mn$_1$ and Mn$_2$ sites, the
Oxygen site would have cubic (O$_{\text{h}}$) symmetry, and the 4 converged WFs
originating from $s$ and $p$ states on the same O$_1$ site would be $sp^3$
hybrids. This is a general result, which is demonstrated in
Appendix~\ref{sec:app_a}: from localized atom--centered $s$--$p$ orbitals, the
maximal localization algorithm always favors $sp^3$--like Wannier functions,
provided that the symmetry is cubic or higher.  This situation occurs,
\textit{e.g.}, in a hypothetical ferromagnetic (FM) MnO crystal, which will be
briefly described later. In AFM MnO, Mn$_1$ and Mn$_2$ sites are
non--equivalent, so that the symmetry of the Oxygen site is reduced to
C$_{3\text{v}}$. However, reminiscence of the $sp^3$ hybrids persists:
disregarding for the moment the $d$ admixture, we see in panel (a) of
Fig.~\ref{fig2} that the WF 1, with $\Gamma_1$ symmetry, displays clearly the
shape of a non--symmetric $p_z$ orbital, with its center shifted from the
atomic site along the [111] direction ($p_z$--$s$ mixing), as also indicated in
Table~\ref{table2}. The three other WFs are equivalent to each other under the
operations of C$_{3\text{v}}$ (they have the same spread), and their centers
are located much closer to the cubic axes than to the diagonals. This can be
explained by considering the interaction with the $3d$ orbitals of Mn$_2$: all
these WFs display a substantial bonding $3d$ contribution from the three
neighboring unoccupied Mn$_2$ sites. As the first--neighbor shell of Mn sites
is an octahedron centered on each O, the above interaction favors the
orientation of the equivalent WFs 2, 3, and 4 along cubic axes. We display in
panel (b) of Fig.~\ref{fig2} the WF 2 whose center is close to the [100] axis.
The sum of WFs centers given in Table~\ref{table2} is a measure of the trigonal
distortion of the $sp^3$--like orbitals, since it would be vanishing in a
perfect cubic environment. This configuration of the converged WFs justifies
also our choice of trial orbitals: $p_x$, $p_y$, and $p_z$ orbitals are already
good guesses of final WFs 2, 3, and 4. During the iterations, most of the
effort is devoted to building up WF 1 (the $p_z$--$s$ hybrid oriented along
[111]). We note also in Fig.~\ref{fig2} small features on the neighboring
Mn$_1$ sites. They correspond to a weak bonding contribution from unoccupied
states on these sites.
\begin{figure}[htbp!]
\includegraphics[scale=0.6]{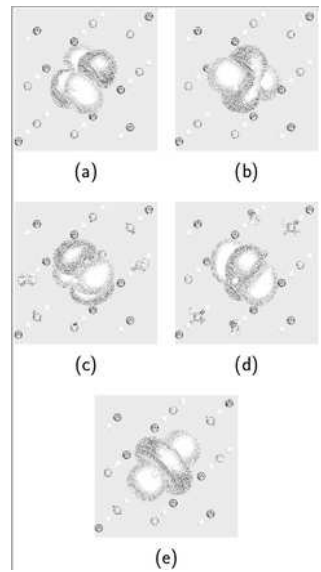}
\caption{Same as Fig.~\ref{fig2}, but for the five Mn$_1$ $3d$ WFs within LSD,
after diagonalization of $\langle r^2 \rangle_{n'n}$. The WFs (a) and (b), as
well as (c) and (d), are partners of a $\Gamma^+_3$ representation of
D$_{3\text{d}}$, while (e) belongs to $\Gamma^+_1$.\label{fig3}}
\end{figure} 

We consider next the five Mn$_1$~$3d$ Wannier functions. They are exactly
centered at the Mn$_1$ site and their spreads do not display any well--defined
symmetry. When convergence of O~$2s$/$2p$ Wannier functions is practically
achieved, further iterations of the localization process have the only effect
of modifying slightly (esentially by amounts comparable to the numerical noise)
the individual spreads of the five Mn$_1$~$3d$ WFs, while their total spread
remains unchanged. These five WFs are not uniquely defined by the criterion of
maximal localization. In fact, any set of WFs obtained by applying a unitary
transformation to the original set, will in general have different individual
spreads, but the same value of the total spread. Since it is advisable that the
final WFs display explicitely as much symmetry as possible, we exploit this
non--uniqueness and perform a symmetry reduction of the 5--fold representation
of the Mn$_1$~$3d$ WFs, which  decomposes into $\Gamma^+_1 + 2\:\Gamma^+_3$
according to D$_{3\text{d}}$ symmetry.

We note that the Hermitian operator $r^2$, which appears in the localization
functional Eq.~(\ref{omega_def}), has the full rotational symmetry, and
therefore its matrix representation $\langle r^2 \rangle_{n'n} = \langle
w_{n'\mathbf{0}} |r^2| w_{n\mathbf{0}} \rangle$ belongs to the $\Gamma^+_1$
representation of D$_{3\text{d}}$. These matrix elements can be calculated
through a simple generalization  of Eq.~(23) of Ref.\ \onlinecite{Marzari97},
and are given by
\begin{eqnarray}
\langle w_{n'\mathbf{0}} |r^2| w_{n\mathbf{0}} \rangle &=& \frac{1}{N}
\sum_{\mathbf{k},\mathbf{b}} w_b\:[2\:\delta_{n'n} -
M_{n'n}^{(\mathbf{k},\mathbf{b})} \nonumber \\*
& & \mbox{} - M_{nn'}^{(\mathbf{k},\mathbf{b})\:*}].
\label{r2_mn}
\end{eqnarray}
We choose the five new WFs as the linear combinations of the original set which
diagonalize the $r^2$ matrix, Eq.~(\ref{r2_mn}). We get in this way an extra
unitary transformation which allows us to update the matrices
$U^{(\mathbf{k})}$. The resulting WFs, which display the expected  ($\Gamma^+_1
+ 2\:\Gamma^+_3$) symmetry, are indicated with the header ``s--LSD'' in
Table~\ref{table3}, and are shown in Fig.~\ref{fig3}. We note that an
arbitrariness (up to a two--by--two unitary transformation) remains in the
definition of the partners in the two $\Gamma^+_3$ representations. The figure
shows that the WFs 9--13 are essentially atomic orbitals modified by the
D$_{3\text{d}}$ crystal field.

\begin{figure}[htbp!]
\includegraphics[scale=0.40]{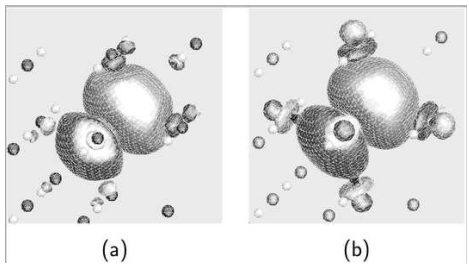}
\caption{LSD Wannier functions of FM MnO. The $sp^3$--like O WF with center
along the [111] direction for (a) majority, and (b) minority--spin. Notations
are the same as in Fig.~\ref{fig2}.\label{fig4}}
\end{figure}

The above picture depicting the whole set of WFs in AFM MnO has been confirmed
by similar calculations performed for a hypothetical FM MnO crystal. In this
case, the symmetry is cubic (space group O$^5_{\text{h}}$), and the
majority--spin chanel contains 18 valence states (O~$2s$, O~$2p$, and Mn~$3d$),
while the minority--spin chanel has 8 valence states (O~$2s$, O~$2p$). As in
AFM MnO, the Mn~$3d$ WFs are found to be essentially atomic orbitals. By
contrast, the O~$2s/2p$ WFs in both chanels are here undistorted $sp^3$ hybrids
(see Appendix~\ref{sec:app_a}). In the minority--spin chanel, \textit{both}
lobes of each hybrid display a substantial bonding $3d$ contribution from the
three neighboring unoccupied Mn sites, while these features are absent in the
majority--spin chanel (occupied Mn sites). The $sp^3$--like Oxygen WFs, with
center along the [111] direction, are shown in Fig.~\ref{fig4} for both spin
chanels.

The covalent interaction in AFM MnO between O and Mn$_2$ $3d$ orbitals, clearly
singled out in Fig.~\ref{fig2} for the case of the spin--up chanel (we remind
that corresponding results for spin--down are obtained by reversing Mn$_1$ and
Mn$_2$), is the fingerprint of superexchange in a one--electron picture.
Superexchange mechanism results in fact from indirect many--body interactions,
involving metal atoms (here Mn$_1$ and Mn$_2$) with an intervening Oxygen. The
shared covalency of nearest--neighbor pairs of magnetic ions leads to an
antiferromagnetic alignment of their moments. As a matter of fact, supposing,
\textit{e.g.} a spin--up WF on the Mn$_1$ site, the neighboring Mn WF has the
choice between two spin orientations, corresponding to a FM or an AFM ordering.
The FM choice will be energetically less favourable, due to the required
orthogonalization to the Mn$_1$ WF, while in the AFM alignment, the spin part
of the wavefunction accounts for this necessary orthogonalization. The above
textbook explanation of AFM ordering can be discussed in light of our
calculations, showing  for the first time a real--space representation of the
WFs for an AFM system. The maximally--localized WFs bring out the relevant
physical information, namely the covalent interaction with empty Mn $3d$
states. We also notice that the FM alignment, on the other hand, is favoured by
kinetic energy. This can be seen in Figs.~\ref{fig2} and \ref{fig4} from the
larger number of covalent interactions (6 instead of 3 in AFM case) associated
with the O $sp^3$--like WFs, which leads eventually to a broader band--width.

\begin{figure}[htbp!]
\includegraphics[scale=0.40,trim=50 0 -50 0,clip]{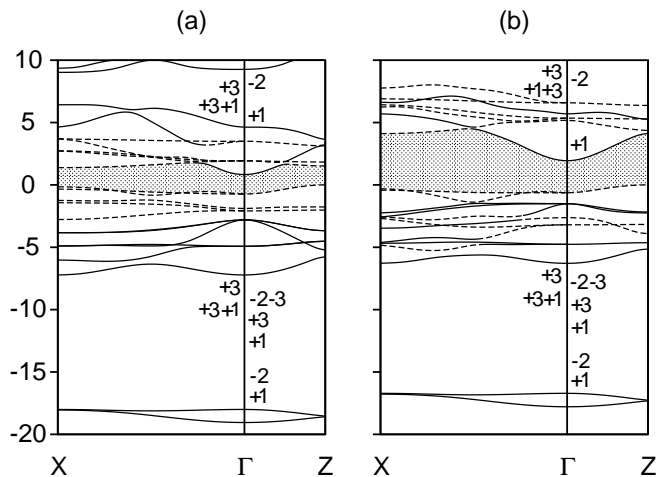}
\caption{Energy bands of antiferromagnetic MnO calculated along the
$X-\Gamma-Z$ lines of the rhombohedral magnetic zone using the
local--spin--density approximation (a), and the LSD + $U$ scheme (b). Shaded
areas correspond to the fundamental gap. Irreducible representations at
$\Gamma$ (Bethe notation) are also given on the left and right of the energy
axis, for states of primary Mn and O origin, respectively. Bands corresponding
to states with high probability in the Mn atomic spheres are drawn with dashed
lines.\label{fig5}}
\end{figure}

It is useful at this point to contrast this rather simple one--electron picture
of AFM MnO, obtained in real space using the localized Wannier--function
description, with the more entangled picture based on the energy
band--structure of the extended Bloch--function description. For this purpose,
we show in the left panel of Fig.~\ref{fig5} the spin--up band structure of MnO
along the $X-\Gamma-Z$ directions, calculated within the LSD scheme.
Starting from the lowest energies, we find first a $\Gamma^+_1$ and a
$\Gamma^-_2$ states at $\mathbf{k}=\mathbf{0}$. They arise from $\Gamma_1$
O~$2s$ orbitals on the two O sites (related by inversion symmetry), forming 
symmetric ($\Gamma^+_1$) and antisymmetric ($\Gamma^-_2$) combinations,
respectively. The primary origin of the next composite group of 6 bands is
O~$2p$ triplets on the two O sites, which are split into a $\Gamma_1$ singlet
and a $\Gamma_3$ doublet by trigonal C$_{3\text{v}}$ site symmetry. They lead
to one set of symmetric combinations ($\Gamma^+_1$, $\Gamma^+_3$), and one set
of antisymmetric combinations ($\Gamma^-_2$, $\Gamma^-_3$) between the two
sites. Next, the occupied $t_{2g}$ orbitals of the Mn$_1$ atom are split at
$\Gamma$ into a doublet plus a singlet ($\Gamma^+_3$, $\Gamma^+_1$), while the
$e_g$ orbitals remain a doublet ($\Gamma^+_3$). These Mn$_1$~$3d$ orbitals give
rise to the uppermost occupied composite group of five bands. Clearly, states
of the Mn$_1$~$3d$ and O~$2p$ complexes which belong to the same $\Gamma^+_1$
(or $\Gamma^+_3$) representation interact with each other, so that eventually
both sets $\Gamma^+_1$(Mn$_1$~$d$), $\Gamma^+_1$(O $p$) and
$\Gamma^+_3$(Mn$_1$~$d$), $\Gamma^+_3$(O $p$) consist of bonding and
antibonding partners of Mn$_1$~$3d$/O~$2p$ hybrids. A mirror picture (relative
to Fermi energy) of the above description applies to the conduction bands which
originate from the Mn$_2$~$3d$ orbitals, if we exclude the lowest conduction
band at $\Gamma$, which is free--electron--like, has mostly Mn $4s$/O $3s$
character, and is not relevant to the present discussion. The above description
would leave completely empty the spin--up $3d$ states of the Mn$_2$ atom.
However, a covalent interaction (clearly depicted in Fig.~\ref{fig2} within the
Wannier--function description) between O~$2p$ and the unoccupied Mn$_2$~$3d$
states takes place, and some amount of  spin--up charge appears at the Mn$_2$
sites, which would be totally empty in a completely ionic picture. For
instance, the highest antibonding $\Gamma^+_3$ valence state at $\Gamma$ has
60\% $d$--occupancy in the Mn$_1$ sphere, 7\% $p$--occupancy in each of the two
O spheres, and 14\% $d$--occupancy in the Mn$_2$ sphere.

\begin{figure}[htbp!]
\includegraphics[scale=0.4]{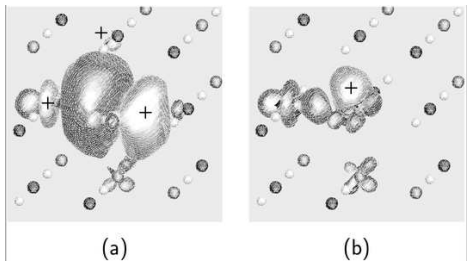}
\caption{(a) The $2s/2p$ WF of O$_1$ as in Fig.~\ref{fig2}~(b), but calculated
within the LSD + $U$ scheme; (b) charge density difference (LSD + $U$ minus
LSD) corresponding to the WF of panel (a). Light grey (a ``+'' sign has been
added for clarity) and dark grey colours indicate positive and negative contour
values, respectively.\label{fig6}}
\end{figure}

With the above results in mind, it is instructive to investigate WFs within
other one--electron schemes, leading to different degrees of wavefunction
localization, covalent interaction, and magnetic moments. Data corresponding to
the LSD + $U$ scheme are presented in Tables~\ref{table2} and \ref{table3} for
the spin--up chanel. Values for the Mn$_1$~$3d$ WFs are those obtained after
diagonalization of $\langle r^2 \rangle_{n'n}$. All individual spreads, as well
as the total spread (whose value is 8.083 \AA$^2$), are smaller than in the LSD
case: the O~$2s$/$2p$ and Mn~$3d$ Wannier functions are more localized by 9\%
and 22\%, respectively, and this is consistent with the narrower valence band
width. This property appears also in Fig.~\ref{fig6}, where we show the LSD +
$U$ result for the WF 2 derived from O~$2s$/$2p$ states.
This LSD + $U$ orbital is similar to the corresponding one calculated within
LSD, and displayed in Fig.~\ref{fig2}b. The most noticeable difference appears
in the weaker covalent interaction between O and the Mn$_2$ $3d$ orbitals. This
is clearly visible in the right panel of Fig.~\ref{fig6}, where we display the
$d$--like differential charge density ($\rho_{\text{LSD}+U} -
\rho_{\text{LSD}}$) associated with this particular Oxygen WF. The five
Mn$_1$~$3d$ WFs, obtained within LSD + $U$, are very similar to the LSD ones,
and are essentially atomic orbitals modified by the crystal field.

The more localized character of all WFs within the LSD + $U$ scheme, together
with the weaker covalent interaction between O and Mn$_2$~$3d$ orbitals, is the
fingerprint in the Wannier--function description of the larger on--site
interaction $U$ of cation $3d$ states. It is worth to realize that this simple
trend is more easily detectable in terms of Wannier functions than Bloch
functions. To this end, we show in the right panel of Fig.~\ref{fig5} the LSD +
$U$ band structure of MnO. It differs in several respects from the LSD one. The
energy gap (2.0 eV), though still smaller than the experimental one ($3.8-4.2$
eV), is closer to it than the LSD value (0.8 eV). The energy separation between
occupied and empty $d$ states, which is due to the on--site interaction $U$
between cation 3$d$ states, has also increased. Furthermore, the three
Mn$_1$~$3d$ bands corresponding to the lower $\Gamma^+_3$ and $\Gamma^+_1$
states ($t_{2g}$) do not form any more a composite group of bands by
themselves, but rather mix with the O~$2p$ complex. This results in a narrowing
of the valence band width (6.3 eV, against 7.2 eV in LSD).

The distinct features of the WFs computed within LSD and LSD + $U$ can be used 
for discussing the behavior of the Born dynamical charge tensor $\mathsf{Z}^*$
within these different many--body approximations. In fact, as $\mathsf{Z}^*$
can be defined in terms of the relative displacement of WFs
centers~\cite{Vanderbilt93}, it carries along a relevant information. In
particular, a $p$--$d$ interaction between valence and conduction states in
mixed ionic--covalent materials~\cite{Posternak97} leads to anomalous
$\mathsf{Z}^*$, while atomic--like occupied WFs correspond to a nominal value.
Therefore in MnO, one expects smaller $|\mathsf{Z}^*|$ (\textit{i.e.} closer to
2, its nominal value) within the LSD + $U$ scheme, to which correspond more
localized WFs, than within LSD. We performed the calculation of the [111]
component of this tensor, using the WF description, and obtained within LSD the
value $|Z^*_{\parallel}| = 2.36$, which is consistent with the one obtained
previously~\cite{Massidda99} using the Berry phase approach. LSD + $U$, on the
other hand, gives as expected a smaller value $|Z^*_{\parallel}| = 2.26$. These
results relate the behavior of two important physical quantities (the
superexchange energy and $\mathsf{Z}^*$) to the same information contained in
the AFM MnO WFs, namely the magnitude of their amplitudes on unoccupied Mn$_2$
sites in the different approximations.

\section{CONCLUSION}
\label{sec:concl}

We have used the maximum localization criterion for constructing Wannier
functions of AFM MnO within both LSD and LSD + $U$ schemes. This is the first
application of the method to a magnetic low--symmetry system, with a partially
filled $d$ band. The LSD + $U$ approach has been considered here, as it
improves the LSD description of the ground--state properties of strongly
correlated materials. The Wannier--function description, which can be viewed as
a generalization to extended systems of the localized molecular--orbitals
description of molecules, provides in the case of MnO a simple picture
complementary to the Bloch--function one. We obtain two distinct groups of WFs:
(i) the five Mn~$3d$ WFs centered on the Mn$_1$ site which are atomic orbitals
modified by the crystal field; (ii) the four trigonally distorted $sp^3$ WFs
associated with each of the O$_1$ and O$_2$ sites, which are not centered on
these sites. The latter display a substantial covalent bonding with $3d$ states
on Mn$_2$ sites, which is consistent with the AFM ordering of this material.
Concerning the Mn WFs, they are not uniquely defined by the criterion of
maximal localization. We have exploited this feature to obtain symmetrized
maximally--localized WFs. We impose this further condition by diagonalization
of the $r^2$ operator.

\begin{acknowledgments}
It is a pleasure to acknowledge stimulating discussions with D. Vanderbilt. We
would like to thank N. Bernstein for providing us with his graphics program
``dan''. This work was supported by the Swiss National Science Foundation
(Grant No. 20--59121.99).
\end{acknowledgments}
 
\appendix
\section{MAXIMALLY LOCALIZED WANNIER FUNCTIONS FROM ONE S--LIKE AND THREE
P--LIKE ORBITALS FOR CUBIC SITE SYMMETRY.}
\label{sec:app_a}

We consider the 4--dimensional Hilbert space ${\mathcal H}_4$ spanned by 4
orthonormal localized basis functions $|\phi_1 \rangle$, $|\phi_2 \rangle$,
$|\phi_3 \rangle$, and $|\phi_4 \rangle$, centered on a given site (taken here
as the origin) with cubic point--group symmetry (T, T$_{\text{h}}$, O,
T$_{\text{d}}$, or O$_{\text{h}}$). $|\phi_1 \rangle$ is the basis function of
a representation $\Gamma_1$ (for T, O, and T$_{\text{d}}$) or $\Gamma^+_1$ (for
T$_{\text{h}}$ and O$_{\text{h}}$). $|\phi_2 \rangle$, $|\phi_3 \rangle$, and
$|\phi_4 \rangle$ are the basis functions of a threefold representation
$\Gamma_4$ (for T and O), $\Gamma_5$ (T$_{\text{d}}$), $\Gamma^+_4$
(T$_{\text{h}}$), or $\Gamma^-_4$ (O$_{\text{h}}$). More precisely, $|\phi_2
\rangle = |\phi_x \rangle$ $\in$ $x$--row of the threefold representation, and
$|\phi_3 \rangle = C^{-1}_3 |\phi_x \rangle$, $|\phi_4 \rangle = C_3 |\phi_x
\rangle$, where $C_3$ is the rotation by $2\pi/3$ around the $[1\;1\;1]$ axis.
We consider real $|\phi_i \rangle$  functions, as the maximally--localized WFs
turn out to be real, apart from an arbitrary overall phase 
factor~\cite{Marzari97}. Finally, we suppose that the $|\phi_i \rangle$ $(i =
1, \ldots, 4)$ are sufficiently localized so that $\langle \phi_i |
\:\mathbf{r}\: | \phi_j \rangle$ and $\langle \phi_i | \:r^2\: | \phi_j
\rangle$ exist and are finite.

We demonstrate below that minimizing within ${\mathcal H}_4$ the localization
functional, Eq.~(\ref{omega_def}), leads to $sp^3$--like Wannier functions, as
opposed to the original atomic--like $|\phi_i \rangle$ functions. The solution
is unique and corresponds to one of the two tetrahedra defined by the cubic
point--symmetry group if the latter is T or T$_{\text{d}}$. For the higher
symmetry groups T$_{\text{h}}$, O, and O$_{\text{h}}$, maximal localization
leads to an infinity of $sp^3$--like solutions constructed from $|\phi_1
\rangle$ and unitary transformed three--fold functions $|\phi_2 \rangle$,
$|\phi_3 \rangle$, $|\phi_4 \rangle$ according to any of the $3\times3$
orthogonal matrices $\mathbf{D}^{(1)} \:(\alpha,\beta,\gamma)$ representing a
rotation of the cartesian co--ordinates of a vector.

As the operator $r^2$ belongs to the identity representation, it is diagonal
in the $|\phi_i \rangle$ basis, and we have
\begin{equation}
\langle \phi_i | \:r^2\: | \phi_i \rangle = \left( 
\begin{array}{cccc}
\Omega_s \; & \; \Omega_p \; & \; \Omega_p \; & \; \Omega_p
\end{array}
\right),
\label{matphir2}
\end{equation}
where cubic symmetry has been taken into account. 
Similarly, writing $\mathbf{r} = x\:\mathbf{i} + y\:\mathbf{j} + z\:\mathbf{k}$
and using projection operators, we find after some algebra
\begin{equation}
\langle \phi_i | \:\mathbf{r}\: | \phi_j \rangle = \left(
\begin{array}{cccc}
0 & \lambda\:\mathbf{i} & \lambda\:\mathbf{j} & \lambda\:\mathbf{k} \\*
\lambda\:\mathbf{i} & 0 & \delta\:\mathbf{k} & \delta\:\mathbf{j} \\*
\lambda\:\mathbf{j} & \delta\:\mathbf{k} & 0 & \delta\:\mathbf{i} \\*
\lambda\:\mathbf{k} & \delta\:\mathbf{j} & \delta\:\mathbf{i} & 0
\end{array}
\right),
\label{matphi}
\end{equation}
with $\lambda = \langle \phi_s |\xi| \phi_{\xi} \rangle$, where $\xi =
x,\;y,\;z$, and $\delta = \langle \phi_x |y| \phi_z \rangle = \langle \phi_y
|z| \phi_x \rangle = \langle \phi_z |x| \phi_y \rangle$. The phase
arbitrariness of the orbitals allows us to set $\lambda \geq 0$. We note that
symmetry requires $\delta=0$ in the case of T$_{\text{h}}$, O, or
O$_{\text{h}}$. The total spread of the original atomic--like $|\phi_i \rangle$
functions is
\begin{equation}
\Omega_{\phi} = \sum_i [ \langle \phi_i | \:r^2\: | \phi_i \rangle -
| \langle \phi_i | \:\mathbf{r}\: | \phi_i \rangle |^2 ] = \Omega_s
+ 3\Omega_p.
\end{equation}
We consider now the four orbitals $|\psi_i \rangle$ obtained from the
$|\phi_j \rangle$ by a general orthogonal transformation $O$, given explicitely
by
\begin{equation}
|\psi_i \rangle = \sum_j O_{ij} \:|\phi_j \rangle \hspace{0.5cm}
(i,\;j=1,\ldots,4).
\label{ortho}
\end{equation}
There are 6 independent matrix elements, and 10 orthonormality conditions for
columns or rows. Using Eqs.~(\ref{matphir2}), (\ref{ortho}) and the
orthonormality conditions, we obtain
\begin{eqnarray}
\sum_i \langle \psi_i | \:r^2\: | \psi_i \rangle &=& \sum_i [\:O^2_{i1} \:
\Omega_s + (O^2_{i2} + O^2_{i3} + O^2_{i4})\: \Omega_p \:]
\nonumber \\*
&=& \sum_i [\:O^2_{i1} \:\Omega_s + \Omega_p \:(1 - O^2_{i1})\:] \nonumber \\*
&=& \Omega_s + 3\Omega_p,
\end{eqnarray}
which verifies the trace invariance property of Eq.~(\ref{matphir2}) under
orthogonal transformations, and gives for the total spread
$\Omega_{\psi}$ of the functions defined by (\ref{ortho}) the inequality
\begin{eqnarray}
\Omega_{\psi} &=& \sum_i \langle \psi_i | \:r^2\: | \psi_i \rangle - \sum_i |
\langle \psi_i | \:\mathbf{r}\: | \psi_i \rangle |^2 \nonumber \\* &=& \Omega_s
+ 3\Omega_p - \sum_i | \langle \psi_i | \:\mathbf{r}\: | \psi_i \rangle |^2
\nonumber \\* &\leq& \Omega_s + 3\Omega_p = \Omega_{\phi}.
\label{opsibar}
\end{eqnarray}
Therefore, the atomic--like $|\phi_i\rangle$ functions correspond to the
maximum value of the spread, and the problem of the total spread  minimization
reduces to minimizing the second term $-\sum_i |\langle \psi_i | \:\mathbf{r}\:
| \psi_i \rangle|^2$ of the localization functional, which can be evaluated
using Eqs.~(\ref{matphi}) and (\ref{ortho}).

We consider next the two particular orthogonal transformations $\widetilde{O}$
and $\widetilde{O'}$ of the $|\phi_i \rangle$ corresponding to the two possible
choices for the four $sp^3$ hybrid orbitals $|\widetilde{\psi}_i \rangle$ and
$|\widetilde{\psi}_i' \rangle$ (written together below 
$|\widetilde{\psi}_i\:^{(\prime)} \rangle$), and given explicitely by
\begin{equation}
\left\{
\begin{array}{c}
|\widetilde{\psi}\:^{(\prime)}_{111} \rangle
= \frac{1}{2} (|\phi_1\rangle \pm
     |\phi_2\rangle \pm |\phi_3\rangle \pm |\phi_4\rangle) \\*
|\widetilde{\psi}\:^{(\prime)}_{1\bar{1}\bar{1}} \rangle 
= \frac{1}{2} (|\phi_1\rangle \pm |\phi_2\rangle \mp |\phi_3\rangle \mp
     |\phi_4\rangle) \\*
|\widetilde{\psi}\:^{(\prime)}_{\bar{1}1\bar{1}} \rangle
= \frac{1}{2} (|\phi_1\rangle \mp |\phi_2\rangle \pm |\phi_3\rangle \mp
     |\phi_4\rangle) \\*
|\widetilde{\psi}\:^{(\prime)}_{\bar{1}\bar{1}1} \rangle 
= \frac{1}{2} (|\phi_1\rangle \mp |\phi_2\rangle \mp |\phi_3\rangle \pm
     |\phi_4\rangle)
\end{array}
\right.,
\label{dirtet}
\end{equation}
where $i=1,\ldots,4=(111),\ldots,(\bar{1}\bar{1}1)$. Using the $\widetilde{O}$
transformation and Eq.~(\ref{matphi}), we get for the $\langle
\widetilde{\psi}_i | \:\mathbf{r}\: | \widetilde{\psi}_j \rangle$ matrix
elements
\begin{widetext}
\begin{equation}
\frac{1}{2} \left(
\begin{array}{@{}cccc@{}}
(\lambda+\delta)(\mathbf{i}+\mathbf{j}+\mathbf{k}) & 
(\lambda-\delta)\:\mathbf{i} &
(\lambda-\delta)\:\mathbf{j} & (\lambda-\delta)\:\mathbf{k} \\*
(\lambda-\delta)\:\mathbf{i} & 
(\lambda+\delta)(\mathbf{i}-\mathbf{j}-\mathbf{k}) &
 -(\lambda-\delta)\:\mathbf{k} & -(\lambda-\delta)\:\mathbf{j} \\*
(\lambda-\delta)\:\mathbf{j} & -(\lambda-\delta)\:\mathbf{k} &
(\lambda+\delta)(-\mathbf{i}+\mathbf{j}-\mathbf{k}) & 
-(\lambda-\delta)\:\mathbf{i} \\*
(\lambda-\delta)\:\mathbf{k} & -(\lambda-\delta)\:\mathbf{j} &
-(\lambda-\delta)\:\mathbf{i} & (\lambda+\delta)(-\mathbf{i}-
\mathbf{j}+\mathbf{k})
\end{array}
\right).
\label{matpsi}
\end{equation}
\end{widetext}
We note that taking the $\widetilde{O'}$ transformation corresponds to changing
$\lambda$ into $-\lambda$ in Eq.~(\ref{matpsi}) for the $\langle
\widetilde{\psi}_i' | \:\mathbf{r}\: | \widetilde{\psi}_j' \rangle$ matrix
elements. Using this latter equation, together with (\ref{opsibar}), we obtain
the total spreads of the $|\widetilde{\psi}_i \rangle$ and
$|\widetilde{\psi}_i' \rangle$ WFs
\begin{eqnarray}
\Omega_{\widetilde{\psi}} &=& \sum_i [ \langle \widetilde{\psi}_i | \:r^2\: |
\widetilde{\psi}_i \rangle - | \langle \widetilde{\psi}_i | \:\mathbf{r}\: |
\widetilde{\psi}_i \rangle |^2 ] \nonumber \\* &=& \Omega_s + 3\Omega_p -
3(\delta+\lambda)^2 \nonumber \\* &=& \Omega_{\phi} - 3(\delta+\lambda)^2 \leq
\Omega_{\phi},
\label{ompsi}
\end{eqnarray}
and
\begin{equation}
\Omega_{\widetilde{\psi}\:'}  = \Omega_{\phi} - 3(\delta-\lambda)^2 \leq
\Omega_{\phi}.
\label{ompsibar}
\end{equation}
Two cases have to be considered according as $\delta$  is zero (point groups
T$_{\text{h}}$, O, O$_{\text{h}}$) or not (point groups T and T$_{\text{d}}$). 

If $\delta=0$, the two spreads $\Omega_{\widetilde{\psi}}$ and
$\Omega_{\widetilde{\psi} \:'}$ are both equal to $\Omega_{\phi} -
3\:\lambda^2$. We demonstrate first that this latter value is the minimum of
the spread, Eq.~(\ref{opsibar}). To this end, using Eqs.~(\ref{matphi}),
(\ref{ortho}), and the orthonormality conditions  $\sum_j O^2_{ij} = 1$ and
$\sum_i O^2_{i1} = 1$, we are lead to minimizing
\begin{eqnarray}
\Omega_{\psi} &=& \Omega_{\phi} - 4\lambda^2 \sum_i O^2_{i1} \:(O^2_{i2} +
O^2_{i3} + O^2_{i4}) \nonumber \\* &=& \Omega_{\phi} - 4\lambda^2 + 4\lambda^2
\sum_i  O^4_{i1},
\end{eqnarray}
with the constraint
\begin{equation}
C = \sum_i O^2_{i1} - 1 = 0.
\end{equation}
The conditions for an extremum of $\Omega_{\psi}$, introducing the Lagrangian
multiplier  $\Lambda$ are
\begin{equation}
\frac{\partial \Omega_{\psi}}{\partial O_{i1}} + \Lambda \frac{\partial C}
{\partial O_{i1}} = 16 \lambda^2 \: O^3_{i1} + 2\Lambda\: O_{i1} = 0,
\end{equation}
from which we get
\begin{equation}
O^2_{11} = O^2_{21} = O^2_{31} = O^2_{41} = \frac{1}{4},
\label{cond_ai}
\end{equation}
and the result
\begin{equation}
\min \Omega_{\psi} = \Omega_{\phi} - 3\lambda^2 = \Omega_{\widetilde{\psi}}
\equiv \Omega_{\widetilde{\psi}\:'} \leq \Omega_{\phi}.
\end{equation}
We demonstrate next that in addition to $\widetilde{O}$ and $\widetilde{O'}$,
there is in fact an \textit{infinity} of orthogonal transformations producing
$|\psi_i \rangle$ orbitals with this same minimum spread value. For this
purpose, we introduce the operator $R\:(\alpha,\beta,\gamma)$ corresponding to
a geometrical rotation specified by its Euler angles, and its
three--dimensional unitary representation
$D^{(1)}\:(\alpha,\beta,\gamma)_{mm'}$, whose basis functions are the spherical
harmonics $Y_{1m}$, with $m = 1,\;0,\;-1$. Consistently with our choice of the
basis functions $|\phi_2 \rangle$, $|\phi_3 \rangle$, and $|\phi_4 \rangle$, we
consider instead the equivalent, real orthogonal representation
$D^{(1)}\:(\alpha,\beta,\gamma)_{\xi\xi'}$ with $\xi,\:\xi' = x,\;y,\;z$, which
is the usual transformation of the  cartesian co--ordinates of a vector. We
then build from this matrix the following orthogonal transformation acting in
the 4--dimensional Hilbert space  ${\mathcal H}_4$
\begin{equation}
\bm{\mathcal{D}}(\alpha,\beta,\gamma) = \left(
\begin{array}{c|c}
1 & \begin{array}{cccc} 0\;\; &  \;\;0\;\;  &
\;\;0 \end{array} \\* \hline
\begin{array}{c} 0 \\* 0 \\* 0 \end{array} &
\;D^{(1)}\:(\alpha,\beta,\gamma)_{\xi\xi'}
\end{array}
\right).
\label{orot}
\end{equation}
We write $|\widehat{\psi}_i \rangle = \sum_k \widetilde{O}_{ik} \:\sum_j
{\mathcal D} (\alpha,\beta,\gamma)_{kj} \:|\phi_j \rangle$, and get after some
algebra, using Eqs.~(\ref{matphi}) and (\ref{ortho})
\begin{equation}
\Omega_{\widehat{\psi}} \equiv \Omega_{\widetilde{\psi}} = \Omega_{\phi} -
3\:\lambda^2 \hspace{1.0cm} \forall \;\alpha, \:\beta, \:\gamma.
\end{equation}
It is important to remember that the $|\phi_i \rangle$ are basis functions of
irreducible representations of a cubic group, and \textit{not} of the full
rotation group. Therefore, applying the transformation (\ref{orot}) on the
original basis functions $| \phi_i \rangle$ does not result in rotated
$\phi_i\;(\mathbf{r})$ functions, \textit{i.e.}, $\phi_i \: (R^{-1}\:
(\alpha,\beta,\gamma)\:\mathbf{r})$, but in linear combinations of the original
basis functions, defining the same Hilbert space ${\mathcal H}_4$, and carrying
the same chemical information as the original functions.

When $\delta \neq 0$ (case of T and T$_{\text{d}}$), the two transformations
$\widetilde{O}$ and $\widetilde{O'}$ produce $|\widetilde{\psi}_i \rangle$ and
$|\widetilde{\psi}_i' \rangle$ orbitals with different total spreads
\begin{equation}
\left. \begin{array}{lll}
\Omega_{\widetilde{\psi}} < \Omega_{\widetilde{\psi}\:'} & \;\;
\text{ if }\delta > 0 & (\lambda\:\delta \geq 0) \\*
\Omega_{\widetilde{\psi}} > \Omega_{\widetilde{\psi}\:'} & \;\;
\text{ if }\delta < 0 & (\lambda\:\delta \leq 0)
\end{array} \; \right\}.
\end{equation}

We prove finally that $\Omega_{\widetilde{\psi}}$ for $\delta > 0$ (or
$\Omega_{\widetilde{\psi}\:'}$ for $\delta < 0$) is a local minimum of the total
spread. To this end, we consider an infinitesimal orthogonal transformation of
the $|\widetilde{\psi}_i \rangle$ (or $|\widetilde{\psi}_i' \rangle$)
\begin{equation}
O(\epsilon) = I + \epsilon \:A + \frac{1}{2} \:\epsilon^2 A^2 + 
{\mathcal O}(\epsilon^3),
\label{inforth} 
\end{equation}
where $A$ is an antisymmetrical operator defined by 6 real parameters. Using
Eq.~(\ref{matpsi}) and applying (\ref{inforth}) to the sets of $sp^3$ hybrid
orbitals $|\widetilde{\psi}_i \rangle$ and $|\widetilde{\psi}_i' \rangle$, we
obtain for the infinitesimal variations of the spreads, up to second order
terms in $\epsilon$, positive semi--definite quadratic forms in the parameters
of the $A$ operator. Hence the correction to $\Omega_{\widetilde{\psi}}$ (or
$\Omega_{\widetilde{\psi}\:'}$) induced by the infinitesimal transformation,
are non--negative. The spread $\Omega_{\widetilde{\psi}}$
($\Omega_{\widetilde{\psi}\:'}$) therefore corresponds for $\delta > 0$
($\delta < 0$) to a local minimum of the localization functional.

\end{document}